\documentclass[twocolumn,prb,floatfix]{revtex4}

\usepackage{graphicx}
\usepackage{dcolumn}
\usepackage{bm}

\begin{document}

\title{Pressure effects on neutral and charged excitons in self-assembled InGaAs/GaAs quantum dots}

\author{Gustavo A. Narvaez}
\affiliation{National Renewable Energy Laboratory, Golden, Colorado 80401}
\author{Gabriel Bester}
\affiliation{National Renewable Energy Laboratory, Golden, Colorado 80401}
\author{Alex Zunger}
\affiliation{National Renewable Energy Laboratory, Golden, Colorado 80401}

\date{\today}

\begin{abstract}
  By combining an atomistic pseudopotential method with the configuration
  interaction approach, we predict the pressure dependence of the binding
  energies of neutral and charged excitons: $X^0$ (neutral monoexciton),
  $X^{-}$ and $X^{+}$ (charged trions), and $XX^0$ (biexciton) in lens-shaped,
  self-assembled In$_{0.6}$Ga$_{0.4}$As/GaAs quantum dots.  We predict that
  (i) with applied pressure the binding energy of $X^0$ and $X^+$ increases
  and that of $X^-$ decreases, whereas the binding energy of $XX^0$ is nearly
  pressure independent. (ii) Correlations have a small effect in the binding
  energy of $X^0$, whereas they largely determine the binding energy of $X^-$,
  $X^+$ and $XX^0$. (iii) Correlations depend weakly on pressure; thus, the
  pressure dependence of the binding energies can be understood within the
  Hartree-Fock approximation and it is controlled by the pressure dependence
  of the direct Coulomb integrals $J$. Our results in (i) can thus be
  explained by noting that holes are more localized than electrons, so the
  Coulomb energies obey $J^{(hh)}>J^{(eh)}>J^{(ee)}$.
\end{abstract}

\maketitle

The energetics of excitons reflects a balance between single-particle energy
levels ${\cal E}^{(e)}$ and ${\cal E}^{(h)}$ of electrons ($e$) and holes
($h$) in the system, and the many-particle carrier-carrier interactions,
resulting from electron-hole Coulomb and exchange
interactions.\cite{knox_book,cho_review_1979,takagahara_PT_1999} The variation of
excitonic energies under pressure naturally reflects the corresponding
variations in single- vs many-particle energies. Of particular interest are
the pressure variations of excitons confined to nanosize dimensions such as in
quantum dots.\cite{chen_JNN_2004,InP_dot.vs.P,meulenberg_PRB_2002,li_JPCM_2001,ma_JAP_2004,manjon_phys.stat.sol_2003,rybchenko_pssb_2003,gamma-X_dots,li_JPCS_1995,itskevich_pssb_1999,williamson_PRB_1998}
Unlike the case of excitons in higher-dimensional systems, where binding and
its pressure dependence reflects mostly many-particle (correlation) effects,
in zero-dimensional (0D) systems where the geometric dimensions are smaller
than the excitonic radius, binding of neutral and charged excitons results from an interesting {\em interplay}
between single-particle and many-particle effects. Here, we use a realistic
description of both single-particle and many-body effects in self-assembled
In$_{0.6}$Ga$_{0.4}$As/GaAs quantum dots, showing how pressure affects the
different components of exciton binding. 
We distinguish the neutral monoexciton $X^0$ (one $e$, one $h$), from the neutral
biexciton $XX^0$ (two $e$, two $h$), positive trion $X^+$ (one $e$, two $h$)
and negative trion $X^-$ (two $e$, one $h$).
While the effect of pressure on $X^0$ has been measured,\cite{ma_JAP_2004,manjon_phys.stat.sol_2003,rybchenko_pssb_2003,gamma-X_dots,li_JPCS_1995,itskevich_pssb_1999}
to the best of our knowledge, the optical spectroscopy of $X^-$, $X^+$ and $XX^0$
under pressure has not yet been reported. For these reason, we provide
definite predictions of the pressure effects.
Each of the $q$-charged excitons has a spectrum of levels $\{\nu\}$, of which
the lowest is termed the ``ground state of $\chi^q$'' ($\chi=X,\,XX$). This
spectrum is usually expressed by expanding the many-body excitonic states
$|\Psi^{(\nu)}(\chi^q)\rangle$ via a set of Slater determinants
$|\Phi(\chi^q)\rangle$. The latter are constructed from single-particle
electron and hole states and accommodate as many carriers as are present in
$\chi^q$. The single-particle states are solutions to the effective
Schr\"odinger equation

\begin{equation}
\label{SP-effective_eq}
\left\{-\frac{1}{2}\nabla^2+V_{ext}({\bf R})+V_{scr}({\bf R})\right\}\psi_i={\cal E}_i\,\psi_i,
\end{equation}

\noindent where $V_{ext}({\bf R})$ is the external (pseudo) potential (due to
the ion-ion or ion-electron interaction) and $V_{scr}({\bf R})$ is the
screening response to such external potentials. The effect of pressure or
strain is encoded in the ion-ion geometry underlying $V_{ext}({\bf R})$. The many-particle
Hamiltonian is

\begin{eqnarray}
\label{multi_exciton}
H&=& \sum_i\,{\cal E}^{(e)}_i\,c^{\dagger}_ic_i-\sum_j\,{\cal
  E}^{(h)}_j\,h^{\dagger}_jh_j \nonumber \\
 & & +\frac{1}{2}\sum_{ijkl}\,J^{(ee)}_{ij;kl}\,c^{\dagger}_ic^{\dagger}_jc_kc_l
  +\frac{1}{2}\sum_{ijkl}\,J^{(hh)}_{ij;kl}\,h^{\dagger}_ih^{\dagger}_jh_kh_l\nonumber \\ 
 & & -\frac{1}{2}\sum_{ijkl}\left[J^{(eh)}_{ij;kl}-K^{(eh)}_{ij;kl}\right]h^{\dagger}_ic^{\dagger}_jc_kh_l,
\end{eqnarray}

\noindent where $c^{\dagger}_i$ ($c_i$) and $h^{\dagger}_j$ ($h_j$) create (destroy)
an electron in the single-particle state $\psi^{(e)}_i$ and a hole in $\psi^{(h)}_j$,
respectively. In Eq. (\ref{multi_exciton}), the Coulomb and electron-hole
exchange matrix elements are given, respectively, by 

\begin{widetext}
\begin{eqnarray}
\label{coul.scattering}
J^{(\mu\mu^{\prime})}_{ij;kl}=\int\int {\rm d}{\bf R}{\rm d}{\bf
  R}^{\prime}\frac{\left[\psi^{(\mu)}_i({\bf
  R})\right]^{*}\left[\psi^{(\mu^{\prime})}_j({\bf
  R}^{\prime})\right]^{*}\left[\psi^{(\mu^{\prime})}_k({\bf
  R}^{\prime})\right]\left[\psi^{(\mu)}_l({\bf R})\right]}{\epsilon({\bf R},{\bf
  R}^{\prime})|{\bf R}-{\bf R}^{\prime}|}, \\
%
%
\label{K_eh}
K^{(eh)}_{ij;kl}=\int\int {\rm d}{\bf R}{\rm d}{\bf
  R}^{\prime}\frac{\left[\psi^{(h)}_i({\bf
  R})\right]^{*}\left[\psi^{(e)}_j({\bf
  R}^{\prime})\right]^{*}\left[\psi^{(e)}_k({\bf R})\right]\left[\psi^{(h)}_l({\bf
  R}^{\prime})\right]}{\epsilon({\bf R},{\bf R}^{\prime})|{\bf R}-{\bf R}^{\prime}|}.
\end{eqnarray}
\end{widetext}

\noindent Here, $\epsilon({\bf R},{\bf R}^{\prime})$ is a phenomenological,
microscopic dielectric function that screens the Coulomb and exchange
interactions, and in this work we have adopted the $\epsilon({\bf R},{\bf
  R}^{\prime})$ proposed by Resta.\cite{resta_PRB_1977}
The diagonal elements
$J^{(\mu\mu^{\prime})}_{ij}=J^{(\mu\mu^{\prime})}_{ij;ji}$ of Eq.
(\ref{coul.scattering}) are the familiar electron-electron
($\mu\mu^{\prime}=ee$), hole-hole ($\mu\mu^{\prime}=hh$) and electron-hole
($\mu\mu^{\prime}=eh$) direct Coulomb integrals. The electron-electron and
hole-hole {\em exchange} integrals are given by $J^{(ee)}_{ij;ij}$ and
$J^{(hh)}_{ij;ij}$, respectively.  Solving the single-particle Eq.
(\ref{SP-effective_eq}) for a given dot yields the wavefunctions $\psi_i$,
which are used to construct the Slater determinants $|\Phi(\chi^q)\rangle$ for
$\chi^q$ and solve the many-particle, configuration
interaction\cite{franceschetti_PRB_1999} (CI) problem [Eq.
(\ref{multi_exciton})]. This gives the total (ground-state) energy
$E_{CI}(\chi^q)$ of exciton $\chi^q$, as well as excitonic excited states.

The binding energy of the excitonic complexes are defined as

\begin{equation}
\label{binding.eqs_CI}
{
\begin{array}{rcl}
\Delta_{CI}(X^0)&=&\big[{\cal E}^{(e)}_0-{\cal E}^{(h)}_0\big]-E_{CI}(X^0) \\
\Delta_{CI}(X^-)&=&\big[{\cal E}^{(e)}_0+E_{CI}(X^0)\big]-E_{CI}(X^-) \\
\Delta_{CI}(X^+)&=&\big[-{\cal E}^{(h)}_0+E_{CI}(X^0)\big]-E_{CI}(X^+) \\
\Delta_{CI}(XX^0)&=&2E_{CI}(X^0)-E_{CI}(XX^0). 
\end{array}
}
\end{equation}

\noindent In a simplified Hartree-Fock approximation and neglecting the
electron-hole exchange $K^{(eh)}_{00;00}$ (whose magnitude is of the order of
a few to hundreds of $\mu{\rm eV}$)\cite{bayer_PRB_2002} we have

\begin{equation}
\label{binding.eqs_HF}
{
\begin{array}{rcl}
\Delta_{HF}(X^0)& = & J^{(eh)}_{00} \\
\Delta_{HF}(X^-)& = & J^{(eh)}_{00}-J^{(ee)}_{00} \\
\Delta_{HF}(X^+)& = & J^{(eh)}_{00}-J^{(hh)}_{00}\\
\Delta_{HF}(XX^0)& = & 2J^{(eh)}_{00}-\big[J^{(ee)}_{00}+J^{(hh)}_{00}\big] \\
& = &\Delta_{HF}(X^-)+\Delta_{HF}(X^+).
\end{array}
}
\end{equation}

\noindent The latter relation establishes a ``sum rule'' for the binding energy of the
biexciton at the Hartree-Fock level. 
The many-body, correlation effects $\delta(\chi^q)$ in the binding energy can be quantified by comparing the full
solutions in Eq. (\ref{binding.eqs_CI}) to the HF ones in Eq.
(\ref{binding.eqs_HF}),

\begin{equation}
\label{HF+delta}
\Delta_{CI}(\chi^q)=\Delta_{HF}(\chi^q)+\delta(\chi^q)
\end{equation}

In this work, we consider a lens-shaped (base diameter $b=252\,${\AA} and
height $h=35\,${\AA}) In$_{\rm 0.6}$Ga$_{\rm 0.4}$As/GaAs quantum dot and
study how the excitonic binding energies $\Delta_{CI}(\chi^{q})$ depend on
pressure (well below the $\Gamma_{1c}-X_{6c}$ crossover)\cite{goni_PRB_1987}.
We then analyze this dependence in terms of the pressure dependence of (i)
Coulomb integrals $J^{(\mu\mu^{\prime})}_{00}$ and (ii) correlation energies
$\delta(\chi^q)$. In Eq.  (\ref{SP-effective_eq}), we use a screened
pseudopotential expressed as a superposition of screened atomic
pseudopotentials

\begin{equation}
\label{pseudo}
V_{ext}({\bf R})+V_{scr}({\bf R})=V_{SO}+\sum_l\sum_{\alpha}v_{\alpha}[{\bf
  R}-{\bf R}^{(\alpha)}_l;{\rm Tr}(\widetilde\varepsilon)],
\end{equation}

\noindent where $V_{SO}$ is a non-local spin-orbit pseudopotential;\cite{williamson_PRB_2000} $v_{\alpha}$
is a screened pseudopotential for atom of type $\alpha$ that depends on
strain; and ${\bf R}^{(\alpha)}_{l}$ is the vector position of atom $n$ of type $\alpha$ after the atomic positions
within the simulation supercell (quantum dot+GaAs-matrix) have been relaxed, using a valence
force field,\cite{williamson_PRB_2000} in order to minimize the elastic energy of
the nanostructure. The explicit dependence of $v_{\alpha}$
on strain transfers to the electronic Hamiltonian the information on atomic
displacements. $v_{\alpha}$ has been fitted to {\em bulk} properties of GaAs and InAs,
including bulk band structures, experimental deformation potentials and
effective masses, as well as LDA-determined band
offsets.\cite{williamson_PRB_2000}
Equation (\ref{SP-effective_eq}) is solved in a basis of linear combination of
Bloch bands $\{u^{(M)}_{n{\bf k}}({\bf R},\widetilde\varepsilon)\}$ with band
index $n$ and wave vector {\bf k} of material $M$ (=GaAs, InAs) strained\cite{wang_PRB_1999} to
$\widetilde\varepsilon$. Thus, 

\begin{equation}
\psi_i({\bf R})=\sum_{M}\sum_{n,{\bf k}}\,C^{\,(i)}_{M;n,{\bf
    k}}\,\left[\frac{1}{\sqrt{N}}\,u^{(M)}_{n,{\bf k}}({\bf R},\widetilde\varepsilon)e^{i{\bf k}\cdot{\bf R}}\right],
\end{equation}

\noindent where $N$ is the number of primary cells in the simulation supercell that
contains the quantum dot and GaAs matrix. The many-body
configuration-interaction (CI) expansion is taken over all the Slater
determinants $|\Phi(\chi^q)\rangle$ generated within a set of 12 electron and
20 hole single-particle, confined states.
  
{\em Direct-Coulomb vs correlation contributions to binding.---}Table
\ref{Table_1} shows the CI-calculated binding energies $\Delta_{CI}(\chi^q)$
as well as its decomposition [Eq. (\ref{HF+delta})] into Hartree-Fock
$\Delta_{HF}(\chi^q)$ and correlation $\delta(\chi^q)$ contributions.  We see
that (i) the binding energy of the neutral monoexciton $X^0$ is constituted
primarily by HF energy with only $6\,\%$ being due to correlation.  This is in
contrast with $X^0$ in {\em bulk} semiconductors where $\delta(X^0)$ dominates
over $\Delta_{HF}(X^0)$. 
(ii) For $X^-$, $X^+$ and $XX^0$ the HF and correlation contributions to
binding are comparable.  Specifically, while $X^-$ is bound
(positive $\Delta$) already in HF, here $X^+$ and $XX^0$ are unbound in
HF, but become bound by correlation. 
(iii) For each excitonic complex, the magnitude of the correlations depends
weakly on pressure.

\begin{widetext}
\begin{table*}
\caption{{\label{Table_1}}Comparison of Hartree-Fock (HF) and many-body
 configuration-interaction (CI) binding energies (in {\rm meV}),  and
 verification of the ``sum rule'' [$\Delta_{HF}(X^-)+\Delta_{HF}(X^+)$] for 
different pressures. For each excitonic complex $\chi^q$, we present the CI
 binding energy $\Delta_{CI}(\chi^q)$ as a sum of
 the Hartree-Fock binding energy $\Delta_{HF}(\chi^q)$ and correlation-energy
 component $\delta(\chi^q)=\Delta_{CI}(\chi^q)-\Delta_{HF}(\chi^q)$.}
\begin{tabular}{l@{\hspace{0.75cm}}r@{\hspace{0.75cm}}r@{\hspace{0.75cm}}r@{\hspace{0.75cm}}r@{\hspace{0.75cm}}r}
\hline\hline
\multicolumn{1}{c}{Quantity} & $0.2\,{\rm GPa}$ & $0.8\,{\rm GPa}$ & $1.3\,{\rm
 GPa}$ & $1.8\,{\rm GPa}$ & $2.4\,{\rm GPa}$ \\ \hline 
$\Delta_{CI}(X^0)=\Delta_{HF}(X^0)+\delta(X^0)$ &  $20.8+1.3$ & $21.1+1.4$ &
$21.3+1.4$ & $21.6+1.5$ & $21.9+1.5$ \\
$\Delta_{CI}(X^-)=\Delta_{HF}(X^-)+\delta(X^-)$  &  $1.2+1.3$ & $1.0+1.4$ &
$0.8+1.5$ & $0.6+1.6$ & $0.5+1.6$ \\
$\Delta_{CI}(X^+)=\Delta_{HF}(X^+)+\delta(X^+)$  & $-1.8+2.4$ & $-1.5+2.3$ &
$-1.2+2.3$ & $-0.9+2.2$ & $-0.7+2.2$ \\
$\Delta_{CI}(XX^0)=\Delta_{HF}(XX^0)+\delta(XX^0)$ & $-0.6+2.0$ & $-0.4+2.0$ &
$-0.3+2.0$ & $-0.3+2.0$ & $-0.2+2.0$ \\
\hline
$\Delta_{CI}(X^-)+\Delta_{CI}(X^+)$ & $3.1$ & $3.2$ & $3.4$ & $3.5$ & $3.6$ \\
Sum rule & $-0.6$ & $-0.5$ & $-0.4$ & $-0.3$ & $-0.2$ \\
\hline\hline
\end{tabular}
\end{table*}
\end{widetext}

\begin{figure}
\includegraphics[width=8.0cm]{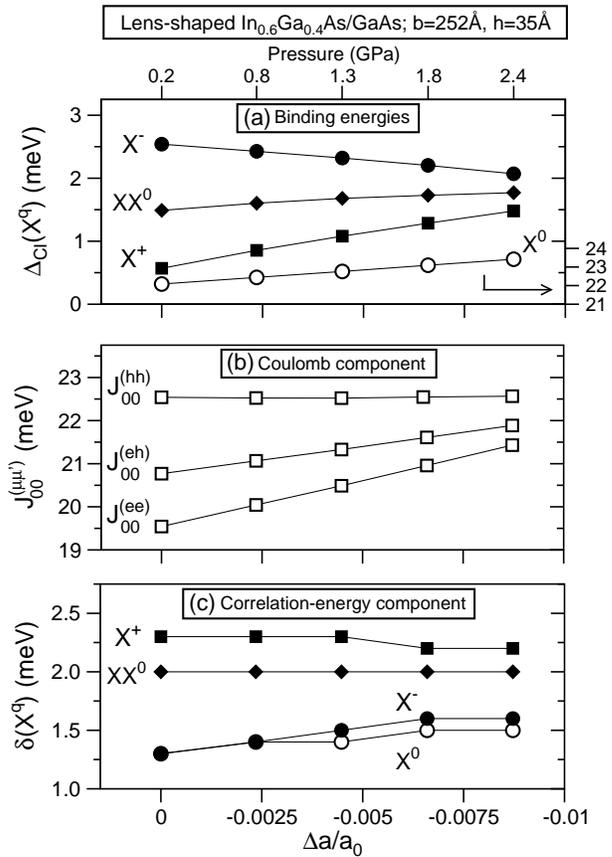}
\caption{{\label{Fig_1}}Pressure dependence of (a) binding energies
  $\Delta_{CI}(\chi^q)$ as obtained from the many-particle,
  configuration-interaction method, (b) Coulomb-energy component
  $J^{(\mu\mu^{\prime})}_{00}$ and (c) correlation-energy components
  $\delta(\chi^q)$. (See text for definitions.)}
\end{figure}

%
{\em Pressure dependence.---}Figure \ref{Fig_1} shows the dependence on
pressure of (a) the binding energies $\Delta_{CI}(\chi^q)$, (b) Coulomb
energies $J^{(ee)}_{00}$, $J^{(eh)}_{00}$ and $J^{(hh)}_{00}$, and (c)
correlation energies $\delta(\chi^q)$. Pressure is represented by $\Delta
a/a_0=(a-a_0)/a_0$, where $a$ and $a_0$ are the distorted and equilibrium
lattice parameter of the GaAs matrix, respectively. The pressure values showed
in the upper axis on Fig. \ref{Fig_1}(a) are calculated by using the equation
of state\cite{welber_PRB_1975}
$P=(B_0/B^{\prime}_0)\big[(V_0/V)^{B^{\prime}_0}-1\big]$, where we take
$V_0/V=[1+{\rm Tr}(\widetilde\varepsilon)]^{-1}$ and calculate ${\rm
  Tr}(\widetilde\varepsilon)$ in the GaAs matrix away from the dot. The
calculation of $\widetilde\varepsilon=\widetilde\varepsilon({\bf R})$ is
performed using atomistic elasticity.\cite{pryor_JAP_1998} We take
$B_0=74.7\,{\rm GPa}$ and $B^{\prime}_0=4.67$ as the GaAs bulk modulus and its
derivative with respect to pressure, respectively.\cite{mcskimin_JAP_1967} We
see from Fig. \ref{Fig_1} that the pressure dependence of the binding energy
of the various excitons is different: (i) $\Delta_{CI}(X^0)$ shows a small,
nearly linear {\em increase} with pressure; changing about $7\,\%$ in the
studied pressure range. $\Delta_{CI}(X^-)$ {\em decreases}
slightly with increasing pressure, while $\Delta_{CI}(X^+)$ {\em increases}
significantly; at $\Delta a/a_0=-0.0087$ ($P=2.4\,{\rm GPa}$) it has increased 
by $160\,\%$ compared to its value at $\Delta a/a_0=0$.  
Similar to the monoexciton case, the binding energy of the biexciton depends
only weakly on pressure, showing a small relative change as pressure reaches $2.4\,{\rm GPa}$.
(ii) Equation (\ref{HF+delta}) shows that the binding has a HF part and
a correlation part. Table \ref{Table_1} showed that the {\em magnitude} of the
binding is decided by the HF part for $X^0$ and by both HF and correlation for
$X^{-}$, $X^{+}$ and $XX^{0}$. However, Fig.  \ref{Fig_1} shows that the {\em
  pressure dependence} is always decided by the HF contribution.
(iii) The ``sum rule'' of Eq. (\ref{binding.eqs_HF}) valid within HF is not
valid at the CI level, quantitatively failing to predict the correct
values of the biexciton binding energies. 
(iv) From (ii), we see that the trends of the binding energies with pressure
are determined by $J^{(eh)}_{00}$, $J^{(ee)}_{00}$ and $J^{(hh)}_{00}$. By
calculating these integrals, we find that
$J^{(hh)}_{00}>J^{(eh)}_{00}>J^{(ee)}_{00}$ and the magnitude of
$J^{(eh)}_{00}$ and $J^{(ee)}_{00}$ increase with a similar slope as pressure
increases while $J^{hh}_{00}$ remains nearly constant; see Fig. \ref{Fig_1}.
This explains the decrease of $\Delta_{CI}(X^-)$ and the increase of
$\Delta_{CI}(X^+)$ with applied pressure. Further, it also becomes clear why
the binding energy of the biexciton remains nearly unchanged with changing
pressure: The similar rate of increase of $J^{(ee)}_{00}$ and $J^{(eh)}_{00}$
with pressure combined with the magnitude of $J^{(hh)}_{00}$ leads to a weakly
pressure-dependent binding energy for $XX^0$.

%

\begin{figure}
\includegraphics[width=8.0cm]{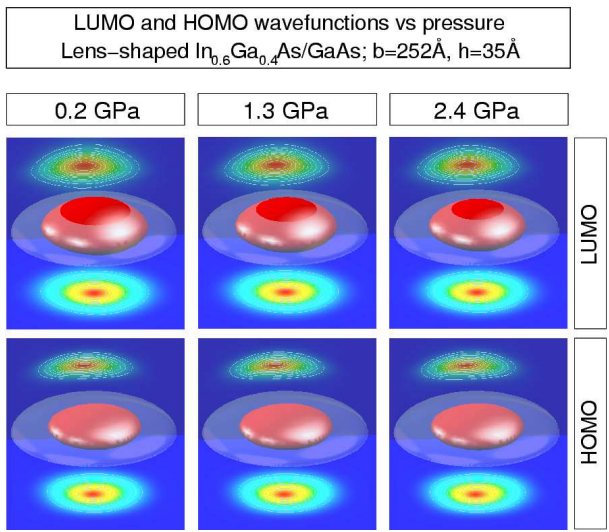}
\caption{{\label{Fig_3}} LUMO $\big[\psi^{(e)}_0\big]$ and HOMO
  $\big[\psi^{(h)}_0\big]$ wavefunctions at $0.2$, $1.3$ and
  $2.4\,{\rm GPa}$. The outline of the dot is present as a light shadow. The
  (red) isosurface encloses 75\% of the charge, the in-plane (bottom) contour
  plot is taken at $1\,{\rm nm}$ above the dot's base and the out-of-plane
  contour plot bisects the dot. The LUMO wavefunction becomes more confined
  as pressure increases---see the how the wavefunction penetrates less into
  the barrier above of the dot as pressure increases; whereas the HOMO
  wavefunctions is almost independent of pressure.}
\end{figure}

%
{\em Wavefunction localization with pressure.---}To understand the trend
$J^{(hh)}_{00}>J^{(eh)}_{00}>J^{(ee)}_{00}$ and also that $J^{(hh)}_{00}$ has
the weakest pressure dependence, while $J^{(ee)}_{00}$ has the strongest,
Figure \ref{Fig_3} shows the calculated wavefunctions for the electron ground
state [$\psi^{(e)}_0$] and hole ground state [$\psi^{(h)}_0$] as a function of
pressure. The isosurfaces enclose 75\% of the charge density, the in-plane
contour plot is taken at $1\,{\rm nm}$ above the base of the dot and the
out-of-plane contour plot bisects the dot.
We see that the electron is always more localized than the hole. In addition,
the electron gets more localized as pressure is applied, while the
localization of the hole remains nearly unchanged. The in-plane (parallel to
the base) spatial extention of the electron does not change as much as the
out-of-plane. In particular, Fig. \ref{Fig_3} clearly shows that the spatial
penetration of the electron wavefunction into the GaAs matrix decreases with
applied pressure. 
The increased localization of the electron with pressure can be explained by
the larger magnitude of the conduction-band edge (CBM) deformation potential
of bulk GaAs with respect to that of bulk InAs,\cite{wei_PRB_1999} which
results in an increased electron confinement in the dot with pressure. In
contrast, the similar magnitude of the valence-band edge (VBM) deformation
potential of both bulk GaAs and InAs leads to small changes in hole
confinement with pressure and, therefore, to small changes in localization.

%

%
{\em Single-particle and excitonic pressure coefficients.---}We calculate the
linear pressure coefficient $a$ by fitting the pressure dependence of the band
gap to $E_g(P)=E_g(0)+a\,P+b\,P^2$. For the dot, at the single-particle (SP)
level, we obtain $a^{\rm (dot)}_{SP}=86.47\,{\rm (meV/GPa)}$, whereas the
excitonic value is $a^{\rm (dot)}_{CI}(X^0)=85.79\,{\rm (meV/GPa)}$. The
latter compares well with the values of $85\,{\rm
  (meV/GPa)}$\cite{ma_JAP_2004}, $80\,{\rm (meV/GPa)}$\cite{gamma-X_dots} and
$82\,{\rm (meV/GPa)}$\cite{gamma-X_dots} observed in InAs/GaAs dots for the
emission lines at $1.28\,{\rm eV}$, $1.26\,{\rm eV}$ and $1.30\,{\rm eV}$,
respectively. For bulk GaAs, we obtain $a^{\rm (bulk)}=105.86\,{\rm
  (meV/GPa)}$, which is within the range of observed values: $94$-$120\,{\rm
  (meV/GPa)}$.\cite{goni_PRB_1987,welber_PRB_1975} Thus, the dot has a smaller
linear pressure coefficient than bulk GaAs.  In addition, it is interesting to
inspect how the VBM and CBM contribute to the linear pressure coefficient of
the band gap. By fitting the lattice-deformation (pressure) dependence of
single-particle eigenvalues ${\cal E}_i(a)={\cal E}_i(a_0)+A_i(\Delta
a/a_0)+B_i(\Delta a/a_0)^2$, we find $A^{\rm (dot)}_{\rm VBM}=-1.76\,{\rm eV}$
and $A^{\rm (dot)}_{{\rm CBM}}=-22.49\,{\rm eV}$. For, bulk GaAs we find
$A_{\rm VBM}=-3.74\,{\rm eV}$ and $A_{\rm CBM}=-27.23\,{\rm eV}$. We see that
the band-gap response to the lattice distortion (pressure) is largely
dominated by the changes in CBM. To reproduce this
LDA-predicted\cite{wei_PRB_1999} behavior, it is necessary to have a
pseudopotential that explicitly depends on strain, otherwise one gets $A_{\rm
  CBM}=-0.46\,{\rm eV}$ and $A_{\rm VBM}=-11.72\,{\rm eV}$ and, consequently,
VBM dominates the gap changes.

%
In summary, we have studied the effects of pressure on the binding energies of
$X^0$, $X^-$, $X^+$ and $XX^0$. Our main findings are the following.  (i) With
applied pressure, the binding energy of $X^0$ and $X^+$ increases and that of
$X^-$ decreases, whereas the binding energy of $XX^0$ is nearly pressure
independent. (ii) The correlation-energy component in the binding energy of
$X^0$ is small, whereas it is large in $X^-$, $X^+$ and $XX^0$; indeed,
correlation is fully responsible for binding the latter complexes. (iii)
Correlations depend weakly on pressure. (iv) The pressure dependence of the
binding energies is controlled by the pressure dependence of the direct
Coulomb integrals. Further, the relative magnitude (order) of these direct
integrals explains the relative magnitude (order) of the binding energies. (v)
Pressure dependence of $J^{(hh)}_{00}$, $J^{(eh)}_{00}$ and $J^{(ee)}_{00}$ is
explained by the changes of the LUMO and HOMO wavefunctions with pressure.

This work was supported by US DOE-SC-BES-DMS.

%

%

%

\begin{thebibliography}{100}

\bibitem{knox_book}R. S. Knox, {\em Theory of Excitons} (Academic Press, New
  York, 1963).

\bibitem{cho_review_1979} K. Cho, in {\em Excitons}, edited by K. Cho (Springer, New
  York, 1979), Chap. 2.

\bibitem{takagahara_PT_1999} T. Takagahara, Phase Transitions {\bf 68}, 281 (1999).

\bibitem{chen_JNN_2004} W. Chen, J. Z. Zhang, and A. G. Joly, J. Nanosci. Nanotech. {\bf 4}, 919 (2004).

\bibitem{InP_dot.vs.P} C. S. Menoni {\em et al.}, Phys. Rev. Lett. {\bf 84}, 4168 (2000).

\bibitem{meulenberg_PRB_2002} R. W. Meulenberg and G. F. Strouse, Phys. Rev. B
  {\bf 66}, 035317 (2002).

\bibitem{li_JPCM_2001} J. Li {\em et al.}, J. Phys.: Condensed Matter {\bf 13}, 2033 (2001).

\bibitem{ma_JAP_2004} B. S. Ma {\em et al.}, J. Appl. Phys. {\bf 95}, 933 (2004)

\bibitem{manjon_phys.stat.sol_2003} F. J. Manj\'on {\em et al.}, phys. stat. sol. (b) {\bf 235}, 496 (2003).

\bibitem{rybchenko_pssb_2003} S. I. Rybchenko {\em et al.}, Phys. Stat. Sol. (b) {\bf 241}, 3257 (2003).

\bibitem{gamma-X_dots}S. G. Lyapin {\em et al.}, Phys. Stat. Sol. (b) {\bf 211}, 79 (1999).

\bibitem{li_JPCS_1995} G. H. Li {\em et al.}, J. Chem. Phys. Solids {\bf 56}, 385 (1995).

\bibitem{itskevich_pssb_1999}I. E. Itskevich {\em et al.}, Phys. Stat. Sol. (b) {\bf 211}, 73 (1999).

\bibitem{williamson_PRB_1998} A. J. Williamson and A. Zunger, Phys. Rev. B {\bf 58}, 6724 (1998).

\bibitem{resta_PRB_1977} R. Resta, Phys. Rev. B {\bf 16}, 2717 (1977).

\bibitem{franceschetti_PRB_1999} A. Franceschetti, H. Fu, L. W. Wang, and
  A. Zunger, Phys. Rev. B {\bf 60}, 1819 (1999).

\bibitem{bayer_PRB_2002}M. Bayer {\em et al.}, Phys. Rev. B {\bf 65} 195315 (2002).

\bibitem{goni_PRB_1987}A. R. Go\~ni, K. Str\"ossner, K. Syassen, and M. Cardona, Phys. Rev. B {\bf 36}, 1581 (1987).

\bibitem{williamson_PRB_2000} A. J. Williamson, L. W. Wang, and A. Zunger,
  Phys. Rev. B {\bf 62}, 12963 (2000).

\bibitem{wang_PRB_1999} L.-W. Wang and A. Zunger, Phys. Rev. B {\bf 59}, 15806 (1999).

\bibitem{welber_PRB_1975}B. Welber, M. Cardona, C. K. Kim, and S. Rodriguez,
  Phys. Rev. B {\bf 12}, 5729 (1975).

\bibitem{pryor_JAP_1998} C. Pryor {\em et al.}, J. Appl. Phys. {\bf 83}, 2548 (1998).

\bibitem{mcskimin_JAP_1967} H. J. McSkimin, A. Jayaraman, and P. Andreatch,
  Jr., J. Appl. Phys. {\bf 38}, 2362 (1967).

\bibitem{wei_PRB_1999}S.-H. Wei and A. Zunger, Phys. Rev. B {\bf 60}, 5404
  (1999).

\end{thebibliography}
\end{document}